\documentclass{elsart}

 \usepackage{epsfig}

\begin{document}

\begin{frontmatter}



\title{Magnetotransport Properties of Antiferromagnetic $YBa_2Cu_3O_{6.25}$ Single Crystals}


\author{  E. Cimpoiasu, and C. C. Almasan}
\address{Department of Physics, Kent State University, Kent  OH 44242}
\author{A. P. Paulikas and B. W. Veal}
\address{Materials Science Division, Argonne National Laboratory, Argonne, IL  60439}
\begin{abstract} 
In-plane $\Delta\rho_{ab}/\rho_{ab}$ and out-of-plane $\Delta\rho_{c}/\rho_{c}$ magnetoresistivities of
antiferromagnetic
$YBa_2Cu_3O_{6.25}$ single crystals were measured in magnetic fields $H$ applied along the
$ab$ plane. $\Delta\rho_{ab}/\rho_{ab}$ is a superposition of two components: The first component is strongly
in-plane anisotropic, changing sign from negative when $H\parallel I$ to positive when $H\perp I$. The second
component is positive, quadratic in $H$, and isotropic in the ab-plane. $\Delta\rho_{c}/\rho_{c}$ displays a
fourfold symmetry upon in-plane rotation of the magnetic field, with maxima along the easy axes of
antiferromagnetic spin ordering and minima along unfavorable directions of spin orientation
($45^{0}$ from the $Cu-O-Cu$ bonds). 

\end{abstract}
\end{frontmatter}

\section{Introduction}

The high $T_c$ cuprates exhibit a large range of behaviors, from antiferromagnetic insulators  to
superconductors and normal metals. The interplay between charge and spin subsystems present in the 
$CuO_2$ planes is considered to be the tuning factor between all these different phases. 
Information about this interaction can be obtained by investigating the charge transport of
the compositions which display antiferromagnetism, the precursor of superconductivity.
Magnetotransport measurements represent an attractive way of
investigation since they can detect changes in the scattering mechanisms of charge carriers
and trace the onset of magnetic ordering. 

Here we present in-plane magnetoresistivity
$\Delta\rho_{ab}/\rho_{ab}=[\rho_{ab}(H)-\rho_{ab}(0)]/ \rho_{ab}(0)$ measurements of antiferromagnetic
$YBa_2Cu_3O_{6.25}$ single crystals, performed in sweeping a magnetic field $H$ applied parallel to the
$ab$ plane and parallel or perpendicular to the electrical current $I$. These measurements
reveal the presence of two terms contributing to $\Delta\rho_{ab}/\rho_{ab}$. The first term is
anisotropic with respect to in-plane magnetic field $H$ orientation, strongly temperature dependent, and
saturates at 
higher fields. The second term is isotropic in the $ab$ plane, positive, and quadratic in $H$. Upon
the in-plane rotation of the magnetic field, the out-of-plane magnetoresistivity $\Delta\rho_{c}/\rho_{c}$
exhibits a fourfold symmetry with maxima along the in-plane crystalographic axes, which are also easy axes of
antiferromagnetic spin-ordering. All these results
are consistent with the presence of antiferromagnetic domains and in-plane orthorhombic distortion of the
crystal lattice due to its coupling to the antiferromagnetically ordered Cu(2) magnetic moments. 

\section{Experimental Details}

Single crystals of antiferromagnetic $YBa_2Cu_3O_{6.25}$ were grown by a method described elsewhere \cite{Veal}.
Typical dimensions are $0.8$x$0.5$x$0.04\;mm^3$ with the c-axis of the single crystals oriented along the smallest
dimension. We have used an eight lead terminal configuration for the simultaneous determination of in-plane
$\rho_{ab}$ and out-of-plane
$\rho_{c}$ resistivities and their respective magnetoresistivities (MR). Thus, we applied a low electrical
current across the top face and measured the top $V_{top}$ and bottom $V_{bot}$ face voltages. The
homogeneity of the single crystals was checked by performing multiple sets of multiterminal transport
measurements  in zero field while sweeping the temperature between $10 \;K\leq T\leq 300 \;K$. The MR
measurements were performed at constant temperature $T$ either by sweeping the magnetic field $H$ up to $14\;T$
or by rotating a magnetic field of $14\;T$ in the $ab$ plane of the single crystal. During magnetic field
sweeps, $H$ was applied parallel to the $ab$ plane of the crystal, with two different orientations with respect
to the electrical current $I$: (i) $H\parallel I \parallel ab$, and (ii) 
$H\perp I \parallel ab$.  Special care was taken to account for the magnetoresistance
of the temperature sensors (Pt or Cx) and to eliminate the contribution of the Hall effect to the measured
voltages \cite{Cimpoiasu}.  

\section{Results and Discussion }

Figure 1 shows the temperature dependence of the in-plane $\rho_{ab}$ and out-of-plane
$\rho_{c}$ resistivities measured in zero
magnetic field on a typical $YBa_{2}Cu_{3}O_{6.25}$ single crystal. This sample is antiferromagnetic for all
measured temperatures ($T
\leq 300\;K$). Note that $\rho_{ab}(T)$ maintains a weak metallic behavior down to
$T \approx 175\;K$, despite the fact that this sample is strongly underdoped. At lower temperatures, $\rho_{ab}(T)$
is nonmetallic, increasing sharply with decreasing $T$. Over the same $T$ range ($10\;K<T<300\;K$), $\rho_{c}(T)$
is very large and displays a nonmetallic behavior.

\begin{figure}
\begin{center}
\epsfig{file=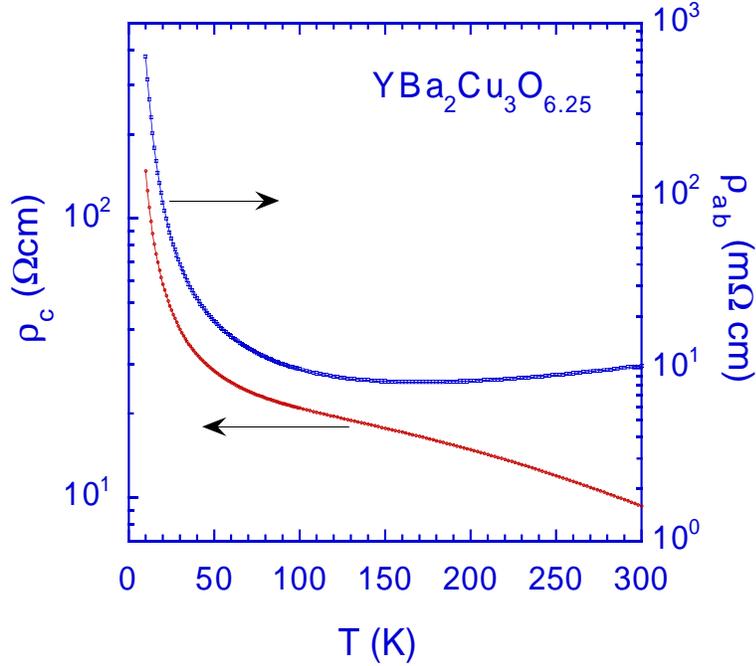,width=4in, height=3.5in}
\end{center}
\caption{Temperature $T$ dependence of in-plane $\rho_{ab}$ and out-of-plane $\rho_{c}$  resistivities of 
$YBa_{2}Cu_{3}O_{6.25}$ single crystal measured in zero magnetic field. }
\end{figure}

The field dependence of the in-plane magnetoresistivity $\Delta\rho_{ab}/\rho_{ab}$ is shown in Fig. 2 for the two
different orientations of
$H$ with respect to $I$ and for several temperatures. The open circles correspond to
data taken with $H\parallel I \parallel ab$, while the squares represent data measured with $H\perp I
\parallel ab$. The main feature of these data is that the field dependence of
$\Delta\rho_{ab}/\rho_{ab}$ is a result of two contributions, each dominant over a different field range: (i) An
anisotropic contribution with respect to field orientation,
$\Delta\rho_{ab,anis}/\rho_{ab}$, dominant at low-$H$ values, which is positive for $H \perp I$ and
negative for $H
\parallel I$. (ii) A contribution $\Delta\rho_{ab,is}/\rho_{ab}$ which is positive and isotropic dominates at
high-$H$ values. 

To better understand the nature of the mechanisms present at this antiferromagnetic concentration, it is useful 
to identify the two contributions to
$\Delta\rho_{ab}/\rho_{ab}$ measured at constant T.  Figure 3(a) shows 
$\Delta\rho_{ab}/\rho_{ab}$ vs H data, measured at $125\;K$ under both field orientations (filled circles).
The small
difference between the high-$H$ curves arises from the small asymmetric contributions of the anisotropic term.
Therefore, one can determine the isotropic term [ square symbols in Fig. 3(a)] by taking the average of
$(\Delta\rho_{ab}/\rho_{ab})(H)$ data measured with the two field orientations (filled circles).  Then, the
anisotropic contribution $\Delta\rho_{ab,anis}/\rho_{ab}$ to $\Delta\rho_{ab}/\rho_{ab}$ is obtained by
subtracting $\Delta\rho_{ab,is}/\rho_{ab}$ term from the measured
$\Delta\rho_{ab}/\rho_{ab}$. The resulting curves for $H \parallel I$ and $H\perp I$ are shown in Fig. 3(a) as
open triangles. These two curves have the same $H$ dependence, namely, $\Delta\rho_{ab,anis}/\rho_{ab}$
varies quadratically with $H$ in low fields and approaches saturation for $H\ge H_{sat}$. Also, the two curves
are not completely symmetric: the anisotropic contribution when 
$H\perp I$ is smaller than the anisotropic contribution when $H\parallel I$. 

\begin{figure}
\begin{center}
\epsfig{file=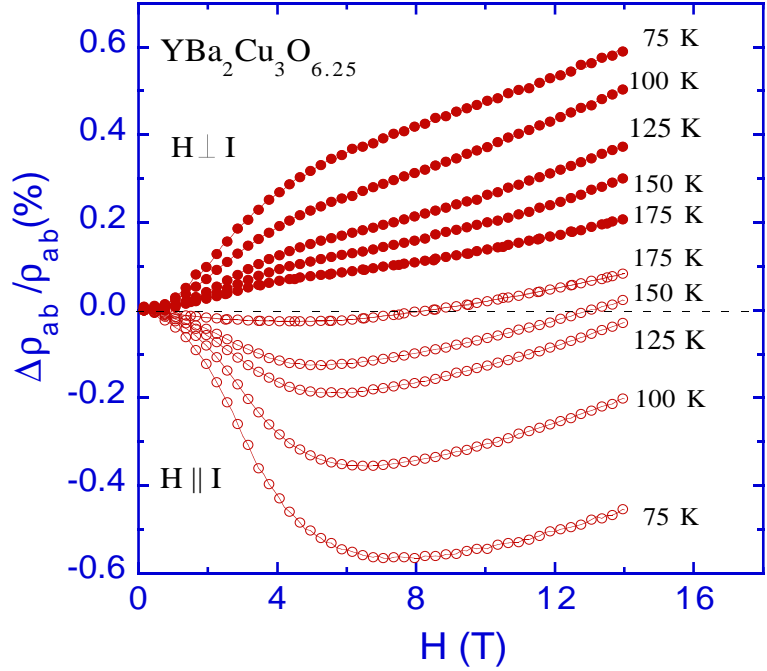,width=4in, height=3.5in}
\end{center}
\caption{ Field $H$ dependence of in-plane magnetoresistivity $\Delta\rho_{ab}/\rho_{ab}$ measured with two
field orientations. The filled circles 
correspond to $H\perp I \parallel ab$ while the open circles correspond to $H\parallel I \parallel ab$. . }
\end{figure}

Figure 3(b) is a plot of the $H$ dependence of $\Delta\rho_{ab,anis}/\rho_{ab}$ scaled with respect to its
value at $H=14\;T$, measured at several $T$ with
$H\perp I$. These data show that for
low-$H$ values, $\Delta\rho_{ab,anis}/\rho_{ab}$ increases faster with $H$ the higher the temperature.
Also, $\Delta\rho_{ab,anis}/\rho_{ab}$ approaches saturation at
$H\geq H_{sat}$. The inset to Fig. 3(b) illustrates the $T$ dependence of $H_{sat}$ and
of $\Delta\rho_{ab,anis}/\rho_{ab}$ measured at $H_{sat}$. Both
$H_{sat}$ and $\Delta\rho_{ab,anis}/\rho_{ab} (H_{sat})$ decrease monotonically with increasing $T$.
Note that $\Delta\rho_{ab,anis}/\rho_{ab}$ changes noticeable with temperature in the
measured $T$ range, in contrast to the isotropic contribution $\Delta\rho_{ab,is}/\rho_{ab}$ which is practically
$T$ independent (see Fig. 2). 

\begin{figure}
\begin{center}
\epsfig{file=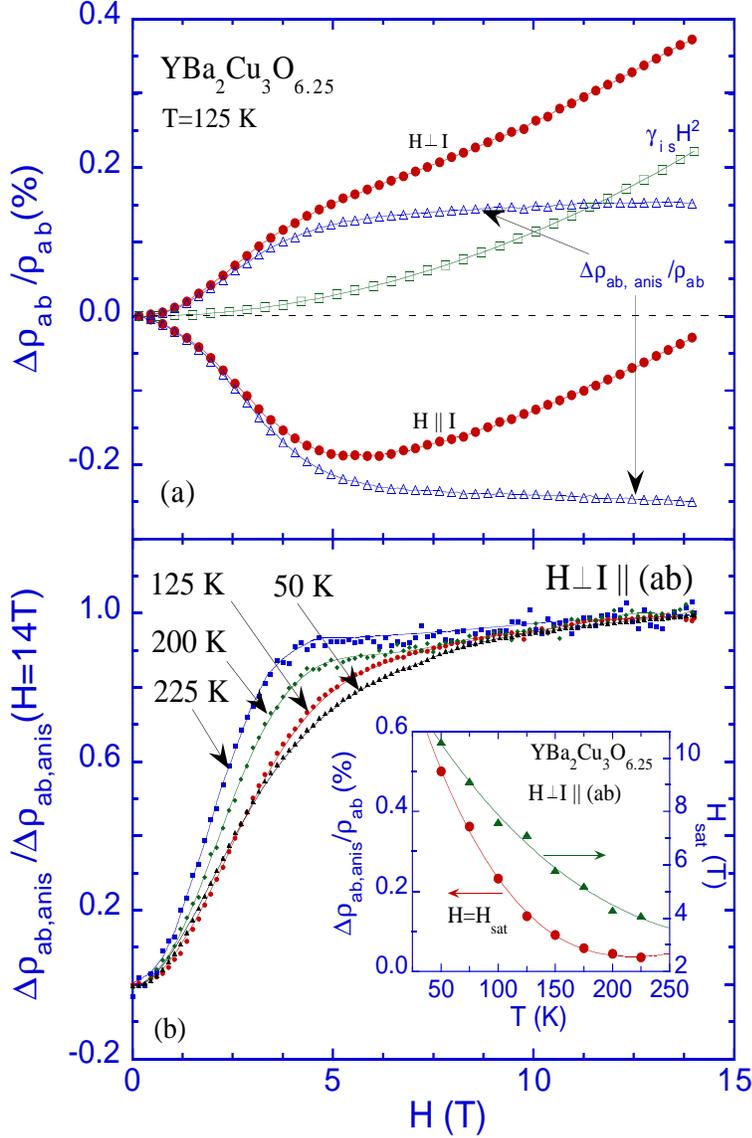,width=4in,height=6in}
\end{center}
\caption{ (a) Magnetic field $H$ dependence of total in-plane magnetoresistivity
$\Delta\rho_{ab, tot}/\rho_{ab}$ measured at $T=125\;K$ with $H\perp I \parallel
ab$ and $H\parallel I \parallel ab$ (filled circles), and of the two terms that contribute to its value:
the anisotropic contribution
$\Delta\rho_{ab,anis}/\rho_{ab}$ (open triangles) and the isotropic $H^2$ contribution
$\Delta\rho_{ab,is}/\rho_{ab}$ (open squares). (b) Magnetic field $H$ dependence of the anisotropic
term $\Delta\rho_{ab,anis}/\rho_{ab}$ for $50\;K \leq T \leq 225\;K$. The data are normalized with respect to
$\Delta\rho_{ab,anis}/\rho_{ab}(H=14\;T)$. Inset: Temperature $T$ dependence of $H_{sat}$ and
of $\Delta\rho_{ab,anis}/\rho_{ab}$ measured at $H_{sat}$.
. }
\end{figure}

Figure 4 is a plot of the out-of-plane magnetoresistivity $\Delta\rho_{c}/\rho_{c}$
measured at
$100\;K$ in a magnetic field of
$14\;T$ vs the angle $\theta$ the field $H$ makes in the $ab$ plane with the crystalographic axis
$a(b)$. Note that
$\Delta\rho_{c}/\rho_{c}$ is anisotropic with a fourfold symmetry. The maxima are at preferred spin
orientations ($0$ and $90^{0}$, hence, along the
$Cu-O-Cu$ bonds), while the minima are at unfavorable spin orientations
($45^{0}$ from the
$Cu-O-Cu$ bonds). This behavior reveals the intrinsic relationship between the out-of-plane conduction and
the antiferromagnetic spin ordering in the $CuO_2$ planes.

\begin{figure}
\begin{center}
\epsfig{file=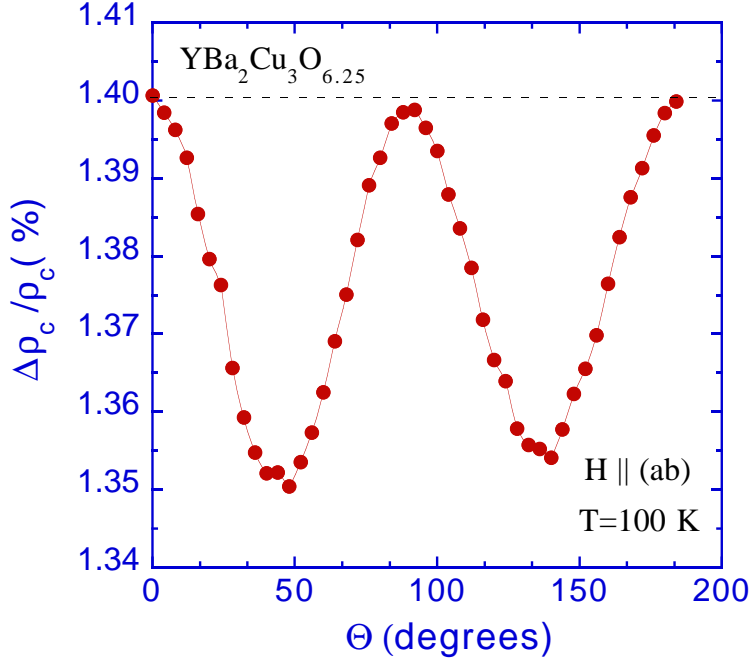,width=4in, height=3.5in}
\end{center}
\caption{ Plot of the out-of-plane magnetoresistivity
$\Delta\rho_{c}/\rho_{c}=[\rho_{c}(\theta)-\rho_{c}(0)]/\rho_{c}$ measured at
$100\;K$ in a magnetic field of
$14\;T$ vs the angle $\theta$ the field $H$ makes in the $ab$ plane with the crystalographic axis
$a(b)$. }
\end{figure}

 Two different
scenarios have been put forward to explain the intriguing behavior of the anisotropic contribution to
$\Delta\rho_{ab}/\rho_{ab}$, which changes sign from positive to negative when
$H$ changes from
$H\perp I$ to $H \parallel I$, respectively. The first scenario invokes  the segregation of charges within an
array of stripes that form the boundaries between the antiferromagnetic domains \cite{Ando}.  The stripes order
along the direction of the magnetic field as a result of the ferromagnetic coupling between the field and the
domain walls. Field ordering of the stripes along $I$ decreases the resistivity from its value
in the absence of a magnetic field, while field ordering of the stripes perpendicular to $I$ increases the
resistivity. As a result,
$\Delta\rho_{ab}/\rho_{ab}<0$ for
$H \parallel I$ and  $\Delta\rho_{ab}/\rho_{ab}>0$ for $H\perp I$. The second scenario invokes the presence of 
an in-plane orthorhombic distortion of the crystal lattice in
the AF state due to its coupling with the ordered Cu(2) magnetic moments
\cite{Janossy1,Janossy2}. This leads to an in-plane anisotropy of the bulk resistivities: resistivity is larger
when $I$ is parallel with the sublattice magnetization and smaller when it is perpendicular. An average
resistivity is measured in the absence of a magnetic field. Since the sublattice magnetization in high fields
is perpendicular to the magnetic field, $\Delta\rho_{ab}/\rho_{ab}<0$ for
$H \parallel I$ and  $\Delta\rho_{ab}/\rho_{ab}>0$ for $H \perp I$.

Thus, both scenarios account for the strong
anisotropy of
$\Delta\rho_{ab}/\rho_{ab}$ and the sign of $\Delta\rho_{ab,anis}/\rho_{ab}$ for the two $H$ orientations with
respect to $I$. However, the anisotropy in $\Delta\rho_{c}/\rho_{c}$ upon field rotation within the
$CuO_2$ plane, with maxima along the in-plane crystalographic axes, is hard to understand in the stripe
scenario unless the stripes also produce a positive $\Delta\rho_{c}/\rho_{c}$ ($H\parallel ab$) which
increases with increasing $H$. According to Ref.
\cite{Ando1} this is not the case, namely, the effect of the magnetic field on the stripes gives rise to a
negative contribution to $\Delta\rho_{c}/\rho_{c}$ for all measured $H$ values. On the other hand, this
angular anisotropy  of $\Delta\rho_{c}/\rho_{c}$, with its fourfold symmetry, can be explained in the
second scenario which invokes the presence of an in-plane orthorhombic distortion of the crystal lattice in
the AF state \cite{Cimpoiasu2}.

None of the two models presented above can explain the positive, isotropic contribution to
$\Delta\rho_{ab}/\rho_{ab}$ when $H\parallel I \parallel ab$ or 
$H\perp I \parallel ab$. A strong, positive, quadratic in $H$ contribution to $\Delta\rho_{c}/\rho_{c}$ when
$H\parallel$ c-axis of samples with similar oxygen concentration was reported to sharply develop upon cooling
through $T_N$ and to persist in the antiferromagnetic state as well
\cite{Ando2}. This contribution was attributed to field suppression of spin fluctuations which might
assist the out-of-plane electron hopping. The change in the ordering state of the spin
subsystem at $T<T_N$ could also affect the in-plane transport
and produce this positive isotropic contribution to $\Delta\rho_{ab}/\rho_{ab}$. 

\section{Summary}

Magnetoresistivity measurements were performed on antiferromagnetic \\
 $YBa_2Cu_3O_{6.25}$ single
crystals in magnetic fields $H \parallel ab$  plane. The in-plane
magnetoresistivity
$\Delta\rho_{ab}/\rho_{ab}$ contains two terms, reflecting two different mechanisms present in this
antiferromagnetic composition. The first term is strongly in-plane anisotropic 
for the two field orientations, $H\parallel I \parallel ab$ and 
$H\perp I \parallel ab$. This term is quadratic in $H$ for low $H$ values
and saturates at higher fields (for example, $H_{sat}=6\;T$ at $T=100\;K$). The second term is positive over 
the whole measured $T$ range, independent of field orientation, and dominates at
$H>H_{sat}$. The out-of-plane resistivity exhibits an in-plane anisotropy, with four-fold symmetry, revealing the
correlation between the out-of-plane transport and preferred directions of antiferromagnetic spin ordering in the
$ab$ plane. All these results are consistent with the presence of antiferromagnetic domains and
an in-plane orthorhombic distortion of the crystal lattice due to its coupling with the ordered
Cu(2) magnetic moments.

{\it This research was supported by the National Science Foundation under Grant No. DMR-9801990 at KSU and the US
Department of Energy under  Contract No. W-31-109-ENG-38 at ANL}.
\label{}

\end{document}